\documentclass[12pt,english,floatfix,nofootinbib,superscriptaddress,aps,prd,preprint]{revtex4}
\usepackage[utf8]{inputenc}
\usepackage{float}
\usepackage{array}
\usepackage{lipsum}
\usepackage{dsfont}
\usepackage{commath}
\usepackage{graphicx}
\usepackage{amsmath,amsthm,amsfonts,amssymb}
\usepackage{graphicx}
\usepackage[english]{babel} 
\usepackage{color}
\usepackage{tensor}
\usepackage{esint}
\usepackage[dvips]{epsfig}
\usepackage[dvips]{graphicx}
\usepackage{float}
\usepackage{units}
\usepackage{textcomp}
\usepackage{mathrsfs}
\usepackage{amsmath}
\usepackage[makeroom]{cancel}
\usepackage{amssymb}
\usepackage{amsbsy}
\usepackage{amsfonts}
\usepackage{amssymb,mathrsfs,xcolor}
\usepackage{esint}
\usepackage{braket}
\usepackage{array}
\usepackage{graphicx}
\usepackage{caption}
\usepackage{subcaption}
\usepackage{comment}

\usepackage{wasysym}
\usepackage{multirow}
\usepackage{wrapfig}

\usepackage{stmaryrd}
\usepackage{upgreek}

\makeatletter

\makeatletter\usepackage{babel}

\usepackage{hyperref}
\hypersetup{
    colorlinks,
    citecolor=blue,
    filecolor=green,
    linkcolor=purple,
    urlcolor=red,
}

\usepackage{slashed}

\newcommand{\ie}{\begin{equation}}
\newcommand{\fe}{\end{equation}}
\newcommand{\se}{\begin{eqnarray}}
\newcommand{\ff}{\end{eqnarray}}

\begin{document}

\title{Quantum gravity effects on particle creation and evaporation in a non--commutative black hole via mass deformation}


\author{A. A. Araújo Filho}
\email{dilto@fisica.ufc.br }

\affiliation{Departamento de Física, Universidade Federal da Paraíba, Caixa Postal 5008, 58051-970, João Pessoa, Paraíba,  Brazil.}
\affiliation{Departamento de Física, Universidade Federal de Campina Grande Caixa Postal 10071, 58429-900 Campina Grande, Paraíba, Brazil}


\author{N. Heidari}
\email{heidari.n@gmail.com}

\affiliation{Center for Theoretical Physics, Khazar University, 41 Mehseti Street, Baku, AZ-1096, Azerbaijan.}

\affiliation{School of Physics, Damghan University, Damghan, 3671641167, Iran.}


\author{Ali \"Ovg\"un}
\email{ali.ovgun@emu.edu.tr}
\affiliation{Physics Department, Eastern Mediterranean University, Famagusta, 99628 North
Cyprus via Mersin 10, Turkiye.}


\date{\today}

\begin{abstract}

In this work, we explore a gravitational non--commutative black hole by gauging the de Sitter SO(4,1) group and employing the Seiberg--Witten map. Specifically, we examine modifications of non--commutativity represented through mass deformation. Initially, we address modifications to \textit{Hawking} radiation for bosonic particle modes by analyzing the Klein--Gordon equation in curved spacetimes. We compute the Bogoliubov coefficients, showing how \(\Theta\) introduces a correction to the amplitude associated with particle creation. Additionally, we derive the power spectrum and the \textit{Hawking} temperature within this framework. We also derive \textit{Hawking} radiation from a tunneling perspective, leading to expressions for the power spectrum and particle number density. A similar analysis is performed for fermion particles. Remarkably, we obtain an analytical expression for black hole evaporation lifetime and compare our results with recent estimates of non--commutativity in the literature.

\end{abstract}

\maketitle


\section{Introduction}

Being a geometric theory of gravity, general relativity is known for its inherently nonlinear nature, which makes finding exact solutions to its field equations particularly challenging, even when additional symmetries and constraints are applied \cite{wald2010general, misner1973gravitation}. To address this complexity, the weak field approximation is often employed. This approach simplifies the field equations significantly, making them more tractable and leading to the prediction of gravitational waves. These waves play a crucial role in the study of stability, \textit{Hawking} radiation from black holes (BHs), and the interactions of BHs with their astrophysical environments.

In general relativity, the framework for describing spacetime geometry does not impose a limit on the precision of distance measurements, a limitation that is thought to be constrained by the Planck length. To tackle this limitation, the concept of non--commutative spacetimes is often introduced. This idea, rooted in string/M--theory \cite{szabo2003quantum,szabo2006symmetry,3}, and has substantial implications in the contexts of supersymmetric Yang--Mills theories \cite{ferrari2004towards,ferrari2003finiteness,ferrari2004superfield}. Additionally, non--commutativity is often introduced into gravitational contexts through the Seiberg--Witten map by gauging a suitable group \cite{chamseddine2001deforming}. Within the non--commutative framework, this approach has led to significant advancements in black hole research \cite{lopez2006towards,nicolini2009noncommutative,mann2011cosmological,modesto2010charged,2,1,zhao2023quasinormal,heidari2024exploring,campos2022quasinormal,karimabadi2020non}, which includes their evaporating features \cite{araujo2023thermodynamics,myung2007thermodynamics} as well as the thermodynamic aspects \cite{lopez2006towards,banerjee2008noncommutative,nozari2006reissner,sharif2011thermodynamics,nozari2007thermodynamics}. Beyond these developments, the thermal properties of field theories have also been explored across various scenarios \cite{araujo2023thermodynamical,furtado2023thermal}.

In modern theoretical physics, the concept of spacetime non--commutativity is described by the relation $[x^\mu, x^\nu] = i \Theta^{\mu \nu}$, where \( x^\mu \) represents spacetime coordinates, and \( \Theta^{\mu \nu} \) is an anti--symmetric tensor. Various approaches have been proposed to integrate non--commutativity into gravitational theories. One method involves using the non--commutative gauge de Sitter (dS) group, SO(4,1), together with the Poincaré group, ISO(3,1), through the Seiberg--Witten (SW) map. Chaichian et al. \cite{chaichian2008corrections} applied this formalism to derive a deformed metric for the Schwarzschild black hole.

In an alternative manner, Nicolini et al. \cite{nicolini2006noncommutative} demonstrated that the effects of non--commutativity could be incorporated into the matter source term of the field equations without modifying the Einstein tensor. This approach involves replacing the point-like mass density on the right--hand side of the Einstein equation with a Gaussian smeared or Lorentzian distribution, specifically \( \rho_\Theta = M (4\pi \Theta)^{-\frac{3}{2}} e^{-\frac{r^2}{4\Theta}} \) or \( \rho_\Theta = M \sqrt{\Theta} \pi^{-\frac{3}{2}} (r^2 + \pi \Theta)^{-2} \), respectively.

Furthermore, Hawking introduced a key idea that connected quantum mechanics with gravity, laying the foundation for quantum gravity theory \cite{o1,o11,o111}. He showed that black holes can emit thermal radiation and gradually evaporate, a process now known as \textit{Hawking} radiation \cite{gibbons1977cosmological,eeeKuang:2017sqa,eeeKuang:2018goo,eeeOvgun:2015box,eeeOvgun:2019jdo,eeeOvgun:2019ygw}. This discovery, based on quantum field theory in curved spacetime near the event horizon, has had a major impact on the study of black hole thermodynamics and quantum effects in strong gravitational fields \cite{o3,o4,araujo2024dark,o6,o7,o8,o9,sedaghatnia2023thermodynamical,araujo2023analysis,aa2024implications}. Afterwards, Kraus and Wilczek \cite{o10}, followed by Parikh and Wilczek \cite{011,o12,o13}, introduced a novel interpretation of \textit{Hawking} radiation as a tunneling process within a semi--classical framework. This approach has since found wide-ranging applications across various black hole models \cite{touati2024quantum,calmet2023quantum,johnson2020hawking,vanzo2011tunnelling,silva2013quantum,anacleto2015quantum,mitra2007hawking,zhang2005new,medved2002radiation,del2024tunneling,mirekhtiary2024tunneling,senjaya2024bocharova}. 

{Furthermore, in Ali et al. (2025), the authors evaluate the physical properties of a Kiselev--like AdS spacetime in the context of $f(R,T)$ gravity under the influence of quantum gravity effects \cite{Ali:2025nrm}. Ali et al. (2024) compute the first--order quantum corrections to tunneling radiation in a modified Schwarzschild--Rindler black hole background \cite{Ali:2024wmj}. Ali et al. (2024) analyze the first-order quantum corrections to the thermodynamics of a dyonic black hole solution surrounded by a perfect fluid \cite{Ali:2024tty}. The authors perform a gravitational analysis of a rotating charged black--hole--like solution within the framework of Einstein--Gauss--Bonnet gravity \cite{Ali:2022omt}. The authors conduct a tunneling analysis incorporating the influences of Einstein--Gauss--Bonnet black hole gravity theory \cite{Ali:2021mtp}.
}

{ This study addresses a gap in the literature by examining particle production in the context of a non--commutative black hole derived from gauge theory. The analysis is motivated by recent developments concerning particle creation mechanisms \cite{calmet2023quantum,araujo2025particle,araujo2025does,filho2025particleasd}, and builds upon a methodology that incorporates mass deformation, as advanced in \cite{heidari2023gravitational,araujo2025neutrino,AraujoFilho:2025viz,AraujoFilho:2024mvz}. The black hole geometry employed here follows the formulation introduced in \cite{Juric:2025kjl}, which corrected and extended the earlier proposal presented in \cite{chaichian2008corrections}, pointing out its limitations in achieving a complete solution. Therefore, the study revisits the phenomenon of Hawking radiation in the setting of a non--commutative black hole background, where both scalar and fermionic modes are taken into account. The Klein--Gordon framework is adopted to explore how the non--commutative parameter $\Theta$ alters the amplitude of particle production, with Bogoliubov coefficients used to extract these modifications. Corrections to the Hawking temperature naturally follow from this procedure. Independently, the tunneling method is applied to analyze particle emission, where the divergent integrals that arise are handled through the residue theorem. The analysis is then extended to fermionic modes, enabling the calculation of their associated particle production rates.

}

\section{Schwarzschild--like black hole via mass deformation } \label{sec2}

The gravitational field deformation was developed by the authors in Ref. \cite{9} through the process of gauging the non--commutative de Sitter SO(4,1) group and applying the Seiberg--Witten map. By contracting the SO(4,1) group to the Poincaré group ISO(3,1), they derived the deformed gravitational gauge potentials, also known as tetrad fields, denoted as \( \hat{e}_\mu^a(x,\Theta) \). These potentials were then utilized in the context of the Schwarzschild black hole, leading to the formulation of a deformed Schwarzschild metric incorporating a non--commutativity parameter up to second order, which is expressed as \cite{Juric:2025kjl}:
{,
\begin{equation}
    \begin{split}
        \hat{g}_{tt}^{\left( r,\Theta \right)} &= -\left(1 - \frac{2 M}{r} \right)+ \frac{M (11 M - 4 r)}{2 r^4}  \Theta^2 + \mathcal{O}(\Theta^3), \\
        \hat{g}_{rr}^{\left( r,\Theta \right)} &= \left(1 -  \frac{2 M}{r}\right)^{-1} + \frac{M (3 M - 2 r) }{2 r^2 (r-2 M)^2} \Theta^2 + \mathcal{O}(\Theta^3), \\
        \hat{g}_{\theta\theta}^{\left( r,\Theta \right)} &= r^2 + \left(\frac{1}{16} -  \frac{2 M}{r} + \frac{M}{8 (r-2 M )}\right) \Theta^2 + \mathcal{O}(\Theta^3), \\
        \hat{g}_{\varphi\varphi}^{\left(r,\Theta ,\theta \right)} &= r^2 \sin ^2\theta + \left(\frac{5 \cos ^2\theta }{16}+\frac{\sin ^2\theta  \left(2 M^2-4 M r+r^2\right)}{4 r (r-2 M)}\right)\Theta ^2 +\mathcal{O}\left(\Theta ^3\right).
    \end{split}
\end{equation}

}

{Although for a direct inspection of the solution $1/\hat{g}_{rr}^{\left( r,\Theta \right)}=0$, there is no corrections ascribed to the non--commutativity on the event horizon \cite{Juric:2025kjl}. Nevertheless, in order to estimate the role of $\Theta$ in our case, we shall consider the expansion of $1/\hat{g}_{rr}^{\left( r,\Theta \right)}$ up to the second order and solve it. In this manner, such an estimation reads} 
\begin{equation}\label{rad}
{r_{s\Theta }} = 2M {-  \frac{\Theta^{2}}{16 M} }.
\end{equation}
The radius  $r_{s\Theta} = 2M_\Theta$ is associated with the deformed non--commutative (NC) mass of the Schwarzschild black hole, leading to the introduction of a newly defined deformed mass, given by:
\cite{heidari2023gravitational}:
\begin{equation}\label{mass}
\ {M_\Theta } = M  {- \frac{\Theta ^2}{32M}}.
\end{equation}
{It is important to mention that a very recent study, Ref.\cite{AraujoFilho:2024mvz}, provided an estimation of the event horizon based on the Chaichian method \cite{chaichian2008corrections}. Nevertheless, the present work is grounded in the approach proposed in Ref.~\cite{Juric:2025kjl}. The difference between both methods in estimating the event horizon, as shown in the previous equation, lies solely in a numerical factor of $1/2$.
}

In this study, the conventional Schwarzschild metric is employed alongside the deformed non--commutative mass described in Eq. (\ref{mass}). 


\section{Boson particle modes}


\subsection{Corrections to \textit{Hawking} radiation}

Initially, let us consider the non--commutative mass deformation of Ref. \cite{heidari2023gravitational}, we obtain
\ie
\label{mainmetric}
\mathrm{d} s^2  = -f(r)\mathrm{d}t^2 + \frac{1}{g(r)}\mathrm{d}r^2 + r^2\mathrm{d}\Omega^2,
\fe    
where 
\ie
f(r) = g(r) =1-\frac{2 M_{\Theta}}{r}. 
\fe

Motivated by these considerations, our aim is to analyze the influence of the quantum parameter $\Theta$ on the process governing the emission of Hawking radiation. The approach adopted here departs from the conventional treatment and focuses on the modifications introduced by non-commutative effects. In the seminal work \cite{hawking1975particle}, Hawking formulated the behavior of a scalar field by studying its wave function within the black hole background, as shown below 
\ie
\frac{1}{\sqrt{-g}}\partial_{\mu}(g^{\mu\nu}\sqrt{-g}\partial_{\nu}\Phi) = 0\label{KL}.
\fe

In the framework of a Schwarzschild geometry, the formulation of the field involves representing the corresponding operator as follows
\begin{eqnarray}
\Phi=\sum_i \left (f_i  a_i+\bar f_i  a^\dagger_i \right)=\sum_i \left (p_i  b_i + \bar p_i  b^\dagger_i + q_i  c_i + \bar q_i  c^\dagger_i \right ) \, .
\end{eqnarray}
The functions $f_i$ and $\bar{f}_i$—with the overbar denoting complex conjugation—are associated with modes that propagate entirely inward. In contrast, $p_i$, $\bar{p}_i$, $q_i$, and $\bar{q}_i$ correspond to outgoing solutions and those lacking any outgoing component. The symbols $a_i$, $b_i$, and $c_i$ stand for annihilation operators, while $a_i^\dagger$, $b_i^\dagger$, and $c_i^\dagger$ are their respective creation counterparts.

In what follows, we aim to establish that the behavior of these mode solutions is subject to modification when quantum gravitational effects—ascribed to the non--commutative parameter $\Theta$—are taken into account. Such corrections to the Schwarzschild background are expected to induce deviations from the standard form of Hawking’s solutions. Consequently, the resulting radiation is no longer the same but instead carries signatures coming from the quantum structure of the spacetime encoded by the non--commutative geometry.

Since both the classical Schwarzschild geometry and its quantum-modified counterpart preserve spherical symmetry, the mode functions—whether ingoing or outgoing—can be expanded in terms of spherical harmonics. In the region outside the event horizon, these solutions admit the following representation \cite{calmet2023quantum}:
\begin{eqnarray}
f_{\omega^\prime l m} &=& \frac{1}{\sqrt{2 \pi \omega^\prime} r }  F_{\omega^\prime}(r) e^{i \omega^\prime v} Y_{lm}(\theta,\phi)\ , \\ 
p_{\omega l m} &=& \frac{1}{\sqrt{2 \pi \omega} r }  P_\omega(r) e^{i \omega u} Y_{lm}(\theta,\phi), 
\end{eqnarray}
The coordinates $v$ and $u$, which correspond to the advanced and retarded time parameters, respectively, take the following forms in the classical Schwarzschild background:  
$v_{\text{classical}} = t + r + 2M \ln \left| \frac{r}{2M} - 1 \right|$ and  
$u_{\text{classical}} = t - r - 2M \ln \left| \frac{r}{2M} - 1 \right|$.  

Using these classical definitions as a starting point, our goal is to extract the dominant quantum correction to these coordinate expressions. A practical route to achieve this is by analyzing the motion of a particle along a geodesic in the given curved spacetime, where the trajectory is parameterized by an affine parameter $\lambda$. Along this path, the momentum of the particle takes the form:
\ie
p_{\mu}=g_{\mu\nu}\frac{\mathrm{d}x}{\mathrm{d}\lambda}^\nu.
\fe
As expected, this momentum remains invariant along the geodesic trajectory, reflecting the conservation laws inherent to the spacetime symmetries. Moreover,
\ie
\label{eq:cons}
\epsilon=-g_{\mu\nu}\frac{\mathrm{d}x}{\mathrm{d}\lambda}^\mu \frac{\mathrm{d}x}{\mathrm{d}\lambda}^\nu .
\fe 

When dealing with massive particles, the parameter choice $\epsilon = 1$ and $\lambda = \tau$—with $\tau$ denoting proper time—ensures consistency with the timelike nature of the path. In contrast, for massless particles, which are the focus of this analysis, one imposes $\epsilon = 0$ and adopts an arbitrary affine parameter $\lambda$.

To proceed, we adopt the general form of a static, spherically symmetric spacetime as given in~\eqref{mainmetric}. Concentrating on radial trajectories, we restrict ourselves to motion with vanishing angular momentum ($p_\varphi = L = 0$), and confine the analysis to the equatorial plane ($\theta = \pi/2$). Under these conditions, the expression for the geodesic evolution simplifies to:
\ie
E =  f(r) \dot{t}.
\fe
Here, the energy $E$ is defined as $E = -p_t$, and the overdot indicates differentiation with respect to the affine parameter $\lambda$, i.e., $\dot{} = \mathrm{d}/\mathrm{d}\lambda$. Following this setup, one also arrives at the relation:
\ie
\label{eq:drdl}
    \left( \frac{\mathrm{d}r}{\mathrm{d}\lambda} \right)^2 = \frac{E^2}{f(r)g(r)^{-1}}.
\fe
After some algebraic manipulations, we have 
\ie
    \frac{\mathrm{d}}{\mathrm{d}\lambda}\left(t\mp r^{*}\right) = 0,
\fe
where $r^*$ is the so--called tortoise coordinate, which reads
\ie
\mathrm{d}r^* = \frac{\mathrm{d}r}{\sqrt{f(r)g(r)}}.
\fe
The advanced and retarded coordinates, $v$ and $u$, respectively, encode the conserved quantities associated with the system. Reformulating the retarded coordinate in terms of the relevant dynamical variables allows us to obtain the following expression:
\ie
\label{lagsd}
\frac{\mathrm{d}u}{\mathrm{d}\lambda}=\frac{2E}{f(r)}.
\fe
For an ingoing geodesic described by the affine parameter $\lambda$, the retarded coordinate $u$ becomes a function of $\lambda$, denoted as $u(\lambda)$. Determining this relation involves two essential procedures: first, one must express the radial coordinate $r$ in terms of $\lambda$; second, the integral presented in \eqref{lagsd} must be evaluated. The resulting form of $u(\lambda)$ important to shaping the Bogoliubov coefficients, which characterize the quantum radiation emitted by the black hole.

To carry out this integration, we utilize the functions $f(r)$ and $g(r)$, integrating the square root in \eqref{eq:drdl} across the domain $r' \in [\tilde{r}_s, r]$, which corresponds to $\lambda' \in [0, \lambda]$. The negative branch of the square root is chosen in this context, as it aligns with the ingoing nature of the geodesic trajectory.

Now, let us assume that quantum deviations remain small in comparison to the dominant classical Schwarzschild terms and considering regions sufficiently near the event horizon, the radial coordinate simplifies to $r = \tilde{r}_s - E\lambda$. Here, $\tilde{r}_s$ denotes the event horizon, which leads to $\tilde{r}_s = 2M_{\Theta}$.

Using this linear approximation for $r(\lambda)$, one can evaluate the integral and obtain the functional form of the retarded coordinate as $u(\lambda) = -4M_{\Theta} \ln\left(\frac{\lambda}{C}\right)$, where $C$ stands as an integration constant.  

Additionally, applying the framework of geometric optics, one establishes a connection between the retarded and advanced null coordinates. This linkage is captured through the relation $\lambda = (v_0 - v)/D$, where $v_0$ corresponds to the value of the advanced coordinate at which the wave is reflected near the horizon (i.e., at $\lambda = 0$), and $D$ is a constant parameter \cite{calmet2023quantum}.

Having established the necessary preliminaries, we now turn to the derivation of the outgoing mode solutions for the modified Klein–Gordon equation that incorporates quantum corrections. The corresponding expressions take the form:
\ie
p_{\omega} =\int_0^\infty \left ( \alpha_{\omega\omega^\prime} f_{\omega^\prime} + \beta_{\omega\omega^\prime} \bar f_{\omega^\prime}  \right)\mathrm{d} \omega^\prime,
\fe
in which $\alpha_{\omega\omega^\prime}$ and $\beta_{\omega\omega^\prime}$ are the
 Bogoliubov coefficients \cite{parker2009quantum,hollands2015quantum,wald1994quantum,fulling1989aspects}
\begin{equation}
\begin{split}
    \alpha_{\omega\omega^\prime}=& -iKe^{i\omega^\prime v_0}e^{\left(2\pi M_{\Theta} \right)\omega} \int_{-\infty}^{0} \,\mathrm{d}x\,\Big(\frac{\omega^\prime}{\omega}\Big)^{1/2}e^{\omega^\prime x}e^{i\omega\Big(4M_{\Theta}\Big)\ln\left(\frac{|x|}{CD}\right)}\\
    = & -i K x \sqrt{\frac{\omega '}{\omega }} e^{2 \pi  M_{\Theta} \omega -i v_{0} \omega '} \left(\frac{x}{\text{CD}}\right)^{4 i M_{\Theta} \omega } \left(-x \omega '\right)^{-4 i M_{\Theta} \omega -1} \Gamma \left(4 i M_{\Theta} \omega +1,-x \omega '\right)
    \end{split}
\end{equation}
and
\begin{equation}
\begin{split}
    \beta_{\omega\omega'} &= iKe^{-i\omega^\prime v_0}e^{-\left (2\pi M_{\Theta}\right)\omega}
    \int_{-\infty}^{0} \,\mathrm{d}x\,\left(\frac{\omega^\prime}{\omega}\right)^{1/2}e^{\omega^\prime x} e^{\left[i\omega\Big(4M_{\Theta} \Big)\ln\left(\frac{|x|}{CD}\right)\right]} \\
    & = -i K x \sqrt{\frac{\omega '}{\omega }} e^{-2 \pi  M_{\Theta} \omega -i v_{0} \omega '} \left(\frac{x}{\text{CD}}\right)^{4 i M_{\Theta} \omega } \left(-x \omega '\right)^{-4 i M_{\Theta} \omega -1} \Gamma \left(4 i M_{\Theta} \omega +1,-x \omega '\right)
    \end{split},
\end{equation}
Here, $\Gamma$ denotes the gamma function, given by the integral representation  
$\Gamma(z) = \int_{0}^{\infty} t^{z-1} e^{-t} \, \mathrm{d}t$.  

This outcome reveals that the amplitude governing quantum particle creation is modified by the corrections introduced through the quantum--altered geometry. In particular, the emergence of non--commutative effects alters the standard amplitude linked to Hawking radiation.

It should be emphasized that, although the quantum gravitational modification alters the amplitude associated with particle production, the resulting power spectrum still retains the form of a blackbody distribution at this level of approximation. To confirm this, it suffices to evaluate:
\begin{equation}\label{eq:alphabetarel}
    |\alpha_{\omega\omega'}|^2 = e^{\big(8\pi M_{\Theta} \big)\omega}|\beta_{\omega\omega'}|^2\,.
\end{equation}
By examining the flux of emitted particles with frequencies within the interval $\omega$ to $\omega + \mathrm{d}\omega$ \cite{o10}, one obtains:
\ie
\label{Fomega}
    P_{\Theta}(\omega)=\frac{\mathrm{d}\omega}{2\pi}\frac{1}{\left \lvert\frac{\alpha_{\omega\omega^\prime}}{\beta_{\omega\omega^\prime}}\right \rvert^2-1}\, ,
\fe
so that we can write
\ie
\label{power}
    P_{\Theta}(\omega)=\frac{\mathrm{d}\omega}{2\pi}\frac{1}{e^{\left(8\pi M_{\Theta}\right)\omega}-1}\,.
\fe
Notice that if we make a comparison with the Planck distribution
\begin{equation} \label{fffgg}
    P_{\Theta}(\omega)=\frac{\mathrm{d}\omega}{2\pi}\frac{1}{e^{\frac{\omega}{T}}-1}
\end{equation}
we, therefore, obtain that
\ie
\label{hawtemp}
    T_{\Theta} = \frac{1}{8\pi M_{\Theta} } \approx \frac{1}{8 \pi  M} {+\frac{\Theta ^2}{256 \pi  M^3}}.
\fe
Up to second order in the non-commutative parameter $\Theta$, the resulting expression aligns with the temperature obtained via the surface gravity method, as established in recent analyses \cite{heidari2023gravitational, araujo2023thermodynamics}. Equation~\eqref{fffgg} shows that a black hole described by a quantum-corrected Schwarzschild geometry emits radiation in a manner consistent with a gray body, characterized by the temperature $T$ given in \eqref{hawtemp}.  {Notice that the first term coincides with the expression obtained for the Schwarzschild black hole, while the second term reflects the non--commutative corrections introduced in the present work. }

It is worth emphasizing, however, that energy conservation across the full system has not yet been incorporated. As radiation is emitted, the black hole’s mass decreases, leading to a corresponding reduction in its size. To derive this dynamical evolution, the next section adopts the tunneling method developed by Parikh and Wilczek \cite{011}, which naturally includes the backreaction due to mass loss.


\subsection{Hawking radiation as a tunneling process}

To account for energy conservation in the emission process from a Schwarzschild black hole modified by quantum corrections through mass deformation, we follow the methodology developed in \cite{011, vanzo2011tunnelling, parikh2004energy, calmet2023quantum}. This framework enables us to compute the radiation spectrum while incorporating backreaction effects due to the black hole’s mass loss.

To facilitate the analysis, the quantum-corrected metric is expressed in Painlevé–Gullstrand coordinates, yielding the line element  
\ie
\mathrm{d}s^2 = -f_{\Theta}(r)\, \mathrm{d}t^2 + 2 h_{\Theta}(r) \, \mathrm{d}t\, \mathrm{d}r + \mathrm{d}r^2 + r^2\, \mathrm{d}\Omega^2,
\fe  
where the cross term is defined by  
\ie
h_{\Theta}(r) = \sqrt{f_{\Theta}(r) \left(g_{\Theta}^{-1}(r) - 1\right)} = \sqrt{\frac{2M_{\Theta}}{r}}.
\fe  

In this context, the tunneling rate is governed by the imaginary part of the classical action, as emphasized in \cite{parikh2004energy, vanzo2011tunnelling, calmet2023quantum}. The action for a particle following a trajectory through the curved background is written as  
\ie
\mathcal{S} = \int p_\mu\, \mathrm{d}x^\mu.
\fe  
When computing $\text{Im} \, \mathcal{S}$, the term involving the radial momentum $p_r\, \mathrm{d}r$ becomes the relevant contribution. The time component $p_t \, \mathrm{d}t = -E\, \mathrm{d}t$ remains purely real and thus does not affect the imaginary part, which governs the tunneling probability
\ie
\text{Im}\,\mathcal{S}=\text{Im}\,\int_{r_i}^{r_f} \,p_r\,\mathrm{d}r=\text{Im}\,\int_{r_i}^{r_f}\int_{0}^{p_r} \,\mathrm{d}p_r'\,\mathrm{d}r.
\fe

It is important to notice that by employing Hamilton’s equations to a system characterized by the Hamiltonian $H = M_{\Theta} - E'$, one finds that $\mathrm{d}H = -\mathrm{d}E'$, where $E'$ ranges from $0$ to $E$, with $E$ denoting the energy carried away by the emitted particle. This relation allows us to write:
\ie
\text{Im}\, \mathcal{S}  = \text{Im}\,\int_{r_i}^{r_f}\int_{M_{\Theta}}^{M_{\Theta}-E} \,\frac{\mathrm{d}H}{\mathrm{d}r/\mathrm{d}t}\,\mathrm{d} r
=\text{Im}\,\int_{r_i}^{r_f}\,\mathrm{d}r\int_{0}^{E} \,-\frac{\mathrm{d}E'}{\mathrm{d}r/\mathrm{d}t}\,.
\fe
Reordering the integration and introducing a suitable variable change, the expression transforms into: $
    \frac{\mathrm{d}r}{\mathrm{d}t} = -h_{\Theta}(r)+\sqrt{f_{\Theta}(r)+h_{\Theta}(r)^2}=1-\sqrt{\frac{\Delta_{\Theta}(r)}{r}}, $
where $\Delta_{\Theta}(r)=(\sqrt{2M_{\Theta}})^2$, we obtain
\ie
\text{Im}\, \mathcal{S} =\text{Im}\,\int_{0}^{E} -\mathrm{d}E'\int_{r_i}^{r_f}\,\frac{\mathrm{d}r}{1-\sqrt{\frac{\Delta_{\Theta}(r,\,E^\prime)}{r}}}.
\fe
Notice that $\Delta_{\Theta}(r)$ becomes dependent on $E'$ due to the substitution of $M_{\Theta}$ with $(M_{\Theta} - E')$ in the modified metric. As a consequence, the integral develops a singularity at the shifted horizon position $r = \tilde{r}_s$, defined by the condition $g(\tilde{r}_s) = 0$. Evaluating the integral by encircling this pole with a counterclockwise contour gives:
\begin{eqnarray}
    \text{Im}\, \mathcal{S}  =-2 \pi  E   (E -2 M_{\Theta})  .
\end{eqnarray}
Following the procedure outlined in \cite{vanzo2011tunnelling}, the emission probability for a Hawking particle, taking into account quantum corrections, is given by:
\ie
\tilde{\Gamma} \sim e^{-2 \, \text{Im}\, S}=e^{-8 \pi  E M_{\Theta}   \left (1-\frac{E}{2 M_{\Theta}}\right)} .
\fe
It is worth emphasizing that in the limit $E \to 0$, the standard thermal spectrum derived by Hawking is restored. Consequently, the radiation spectrum takes the form:
\begin{figure}
    \centering
      \includegraphics[scale=0.6]{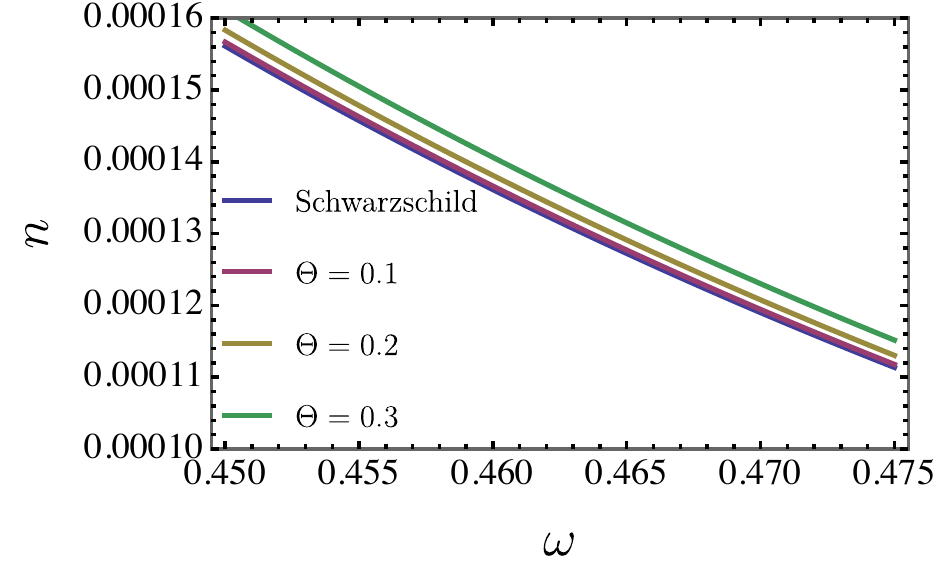}
    \caption{The representation of the particle density for bosons as function of $\omega$ for different values of $\Theta$.}
    \label{particledensity}
\end{figure}
\begin{equation}
    P_{\Theta}(\omega)=\frac{\mathrm{d}\omega}{2\pi}\frac{1}{e^{8 \pi  \omega M_{\Theta}   \left (1-\frac{\omega}{2 M_{\Theta}}\right)
    }-1}.
\end{equation}
Owing to the presence of an extra $\omega$--dependence, the resulting emission spectrum no longer matches the conventional blackbody form, as can be directly confirmed. In the low--frequency regime, however, the expression reduces to a Planck-like distribution characterized by the corrected Hawking temperature. Moreover, the tunneling probability provides a basis for defining the particle number density as follows:
\ie
n = \frac{\tilde{\Gamma}}{1 - \tilde{\Gamma}} = \frac{1}{e^{{-\frac{\pi  \omega  \left(\Theta ^2+16 M (\omega -2 M)\right)}{4 M}} }-1}.
\fe
{ To provide a clearer interpretation of $n$, Fig.~\ref{particledensity} illustrates its dependence on the non--commutative parameter $\Theta$. As expected, the non--commutative corrections introduce small modifications to the particle number density. Furthermore, increasing $\Theta$ leads to an increment in the particle production rate, particularly for low--frequency modes.

}

These results suggest that black hole radiation encodes details about the internal quantum structure of the system. The amplitudes associated with Hawking radiation are modified by the presence of the non-commutative parameter $\Theta$, and the resulting spectrum departs from the ideal blackbody form once quantum gravitational corrections and energy conservation are taken into account.  

Finally, in Tab. \ref{radiation}, we compare our results with recent literature. Recently, similar analyses have been conducted using the deformation of the metric itself \cite{touati2024quantum}. In this manner, our study fills a gap in the literature by incorporating mass deformation in addition to modifications of the metric.

{ Moreover, the non--commutative extension of the AdS group $\text{SO}(4,1)$ provides the foundation for constructing black hole solutions within the gauge theory of gravity in non--commutative spacetime \cite{Juric:2025kjl}. Through the Moyal product, the underlying algebra is deformed, leading to modifications in the geometric structure of spacetime. This procedure introduces corrections to all components of the metric, including the angular sector, which is no longer preserved in general. The resulting non--commutative black hole configurations are still compatible with the semiclassical tunneling framework, as long as the corrections are properly implemented in the near--horizon analysis. Although such an analysis is indeed relevant and deserves attention, it lies beyond the scope of the present work and will be addressed in a forthcoming paper.

}

\begin{table}[t]
	\centering
		\caption{\label{radiation} Comparison of the quantum emission rate, \(\tilde{\Gamma}\), and particle number density, \(n\), for two different approaches: mass deformation and metric deformation.}
	\setlength{\arrayrulewidth}{0.3mm}
	\setlength{\tabcolsep}{30pt}
	\renewcommand{\arraystretch}{1}
	\begin{tabular}{c c c}
		\hline \hline
		~ & Deformed mass (our case) & Deformed metric \cite{touati2024quantum}\\ \hline
		$\Tilde{\Gamma}$ & $e^{-8 \pi  \omega M_{\Theta}   \left (1-\frac{\omega}{2 M_{\Theta}}\right)}$ & $ e^{-2\pi\omega \left( 4M + \frac{3 \Theta^{2}}{2M} \right) }$ \\
		$n$ & $\frac{1}{e^{{-\frac{\pi  \omega  \left(\Theta ^2+16 M (\omega -2 M)\right)}{4 M}} }-1}$ &$\frac{1}{e^{8\pi M \omega \left(1 + \frac{3\Theta^{2}}{8M^{2}} \right)}  -1} $  
  \\ \hline\hline
	\end{tabular}
\end{table}


\subsection{Fermion particle modes}

Since black holes exhibit a well--defined temperature, they emit radiation similarly to an ideal blackbody, aside from modifications introduced by greybody factors. This emission is expected to include particles of various spins, such as fermions. In a series of investigations beginning with the work of Kerner and Mann \cite{o69} and followed by subsequent analyses \cite{o70,o71,o72,o73,o74,o75}, it was shown that both bosonic and fermionic particles, regardless of mass, are emitted at the same thermal temperature. Moreover, even when focusing on spin-1 bosons, it was found that the Hawking temperature remains unchanged under the influence of higher--order quantum corrections \cite{o76,o77}, confirming the robustness of this thermal feature across different particle types and quantum refinements.

In the case of fermionic particles, the action is commonly associated with the phase of the spinor wave function, which satisfies the Hamilton–Jacobi equation. An alternative formulation expresses the fermionic action as \cite{o83,o84,vanzo2011tunnelling}  
\ie
I_f = I_0 + \text{(spin-dependent corrections)},
\fe 
where $I_0$ corresponds to the classical scalar particle action. The additional terms account for the coupling between the particle’s spin and the spin connection of the background spacetime. These contributions, however, do not produce singularities at the event horizon and are typically small in magnitude. Their primary role lies in affecting spin precession, which has negligible impact in this context and can therefore be safely omitted.

Furthermore, the influence of emitted particle spin on the black hole’s total angular momentum is considered insignificant—especially for non-rotating black holes and those with masses well above the Planck scale \cite{vanzo2011tunnelling}. On statistical grounds, emissions of particles with opposite spin orientations occur symmetrically, resulting in no net transfer of angular momentum to or from the black hole.

Expanding upon the framework previously established, we now focus on the quantum tunneling of fermionic particles through the event horizon of the black hole under investigation. The emission rate is computed within a coordinate system analogous to the Schwarzschild geometry. For studies conducted in alternative coordinate systems, such as the generalized Painlevé–Gullstrand or Kruskal–Szekeres frameworks, readers are referred to the foundational analysis in \cite{o69}.

To set the stage, we consider a general spherically symmetric line element of the form:  
\ie
\mathrm{d}s^{2} = {-} \tilde{A}(r) \, \mathrm{d}t^{2} + \frac{1}{B(r)} \, \mathrm{d}r^{2} + \tilde{C}(r) \left( \mathrm{d}\theta^{2} + r^{2} \sin^{2}\theta \, \mathrm{d}\varphi^{2} \right).
\fe

In curved spacetime, the dynamics of spin-$\tfrac{1}{2}$ particles are governed by the Dirac equation, given by  
\ie
\left(\gamma^\mu \nabla_\mu + \frac{m}{\hbar}\right) \Psi(t, r, \theta, \varphi) = 0,
\fe  
where the covariant derivative is expressed as  
\ie
\nabla_\mu = \partial_\mu + \frac{i}{2} \Gamma^{\alpha}_{\;\mu}{}^{\beta} \, \Xi_{\alpha\beta},
\qquad \text{with} \qquad 
\Xi_{\alpha\beta} = \frac{i}{4} [\gamma_\alpha, \gamma_\beta].
\fe

\begin{figure}
    \centering
      \includegraphics[scale=0.6]{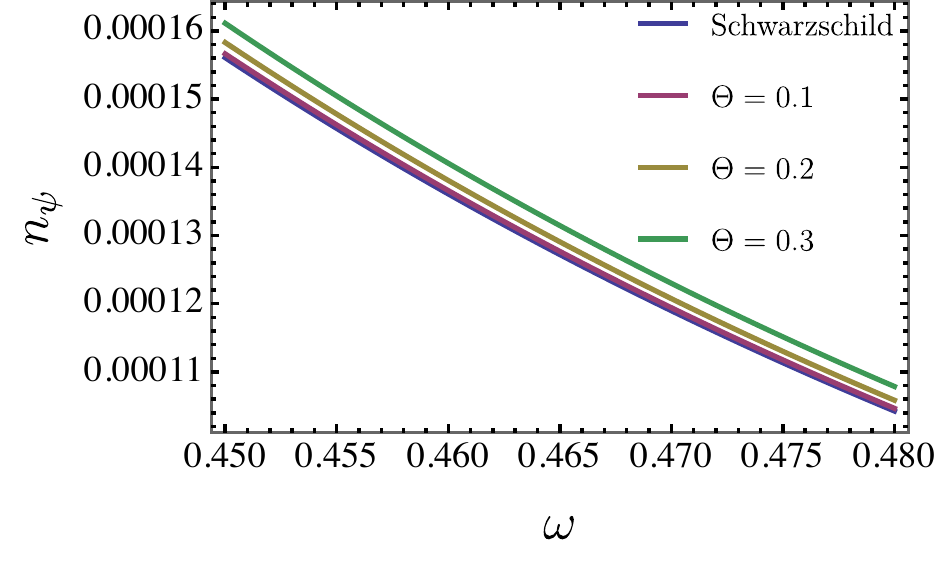}
    \caption{The representation of the particle density for fermions as function of $\omega$ for different values of $\Theta$.}
    \label{particledendityfermions}
\end{figure}

The matrices $\gamma^\mu$ obey the Clifford algebra, satisfying the anticommutation relation  
\ie
\{\gamma_\alpha, \gamma_\beta\} = 2 g_{\alpha\beta} \, \mathbb{I},
\fe  
where $\mathbb{I}$ denotes the $4 \times 4$ identity matrix. For the analysis to follow, we choose an explicit representation for the $\gamma$ matrices suitable to the metric under consideration
\begin{eqnarray*}
 \gamma ^{t} &=&\frac{i}{\sqrt{\tilde{A}(r)}}\left( \begin{array}{cc}
\vec{1}& \vec{ 0} \\ 
\vec{ 0} & -\vec{ 1}%
\end{array}%
\right) \;\;
\gamma ^{r} =\sqrt{\tilde{B}(r)}\left( 
\begin{array}{cc}
\vec{0} &  \vec{\sigma} ^{3} \\ 
 \vec{\sigma} ^{3} & \vec{0}%
\end{array}%
\right) \\
\gamma ^{\theta } &=&\frac{1}{r}\left( 
\begin{array}{cc}
\vec{0} &  \vec{\sigma} ^{1} \\ 
 \vec{\sigma} ^{1} & \vec{0}%
\end{array}%
\right) \;\;
\gamma ^{\varphi } =\frac{1}{r\sin \theta }\left( 
\begin{array}{cc}
\vec{0} &  \vec{\sigma} ^{2} \\ 
 \vec{\sigma} ^{2} & \vec{0}%
\end{array}%
\right)
\end{eqnarray*}%
where the \(\vec{\sigma}\) matrices are the Pauli matrices, which satisfy the standard commutation relations:
$
 \sigma_i  \sigma_j = \vec{1} \delta_{ij} + i \varepsilon_{ijk} \sigma_k, \,\, \text{in which}\,\, i,j,k =1,2,3\;. 
$ The matrix for $\gamma^{5}$ is instead
\begin{equation*}
\gamma ^{5}=i\gamma ^{t}\gamma ^{r}\gamma ^{\theta }\gamma ^{\varphi }=i\sqrt{\frac{\tilde{B}(r)}{\tilde{A}(r)}}\frac{1}{r^{2}\sin \theta }\left( 
\begin{array}{cc}
\vec{ 0} & - \vec{ 1} \\ 
\vec{ 1} & \vec{ 0}%
\end{array}%
\right)\:.
\end{equation*}
For the spin--up Dirac field oriented in the positive \(r\)--direction, we use the following ansatz \cite{vagnozzi2022horizon}:
\begin{equation}
\Psi _{+ }(t,r,\theta ,\varphi ) = \left( \begin{array}{c}
\mathcal{H}(t,r,\theta ,\varphi ) \\ 
0 \\ 
\mathcal{Y}(t,r,\theta ,\varphi ) \\ 
0%
\end{array}%
\right) \exp \left[ \frac{i}{\hbar }\psi_{+}(t,r,\theta ,\varphi )\right]\;.
\label{spinupbh} 
\end{equation} 
We will address the spin--up case explicitly, as the spin--down, which accounts for the negative \(r\)--direction, case is analogous. Inserting the ansatz (\ref{spinupbh}) into the Dirac equation produces:
\ie
\begin{split}
-\left( \frac{i \,\mathcal{H}}{\sqrt{\tilde{A}(r)}}\,\partial _{t} \psi_{+} + \mathcal{Y} \sqrt{\tilde{B}(r)} \,\partial _{r} \psi_{+}\right) + H {i}m &=0, \\
-\frac{\mathcal{Y}}{r}\left( \partial _{\theta }\psi_{+} +\frac{i}{\sin \theta } \, \partial _{\varphi }\psi_{+}\right) &= 0, \\
\left( \frac{i \,\mathcal{Y}}{\sqrt{\tilde{A}(r)}}\,\partial _{t}\psi_{+} - \mathcal{H} \sqrt{\tilde{B}(r)}\,\partial_{r}\psi_{+}\right) + \mathcal{Y} {i}m &= 0, \\
-\frac{\mathcal{H}}{r}\left(\partial _{\theta }\psi_{+} + \frac{i}{\sin \theta }\,\partial _{\varphi }\psi_{+}\right) &= 0,
\end{split}
\fe%
for the leading order in $\hbar$. We assume the action has the form
$
\psi_{+}=- \omega\, t + \chi(r) + L(\theta ,\varphi )  $
which produce the following equations
\cite{vanzo2011tunnelling}
\begin{eqnarray}
\left( \frac{i\, \omega\, \mathcal{H}}{\sqrt{\tilde{A}(r)}} - \mathcal{Y} \sqrt{\tilde{B}(r)}\, \mathcal \chi^{\prime }(r)\right) +m\,{i} \mathcal{H} &=&0,
\label{bhspin5} \\
-\frac{\mathcal{H}}{r}\left( L_{\theta }+\frac{i}{\sin \theta }L_{\varphi }\right) &=&0,
\label{bhspin6} \\
-\left( \frac{i\,\omega\, \mathcal{Y}}{\sqrt{\tilde{A}(r)}} + \mathcal{H}\sqrt{\tilde{B}(r)}\, \mathcal \chi^{\prime }(r)\right) +\mathcal{Y}\,{i}m &=&0,
\label{bhspin7} \\
-\frac{{\mathcal{Y}}}{r}\left( L_{\theta } + \frac{i}{\sin \theta }L_{\varphi }\right) &=& 0.
\label{bhspin8}
\end{eqnarray}
Regardless of the specific forms of \(\mathcal{H}\) and \(\mathcal{Y}\), Equations (\ref{bhspin6}) and (\ref{bhspin8}) yield \(L_{\theta} + i(\sin \theta)^{-1} L_{\varphi} = 0\), indicating that \(L(\theta, \varphi)\) must be a complex function. This solution for \(L\) holds for both outgoing and incoming cases. As a result, when the outgoing probability is divided by the incoming probability, the contribution from \(L\) cancels out, allowing us to disregard \(L\) moving forward. Thereby, Eqs. (\ref{bhspin5}) and (\ref{bhspin7}), for the massless case, present two possible solutions $
\mathcal{H} = -i \mathcal{Y} , \qquad \mathcal \chi^{\prime }(r) \equiv \chi_{ out}'= \frac{\omega}{\sqrt{\tilde{A}(r)\tilde{B}(r)}}, \,\,\,\,\,\,\,
\mathcal{H} = i  \mathcal{Y} ,\qquad  \chi^{\prime }(r)\equiv \mathcal \chi_{ in}^{\prime }(r)= - \frac{\omega}{\sqrt{\tilde{A}(r)\tilde{B}(r)}}$
with $ \chi_{out}$ and $ \chi_{in}$ represents outward and inward solutions, respectively \cite{vanzo2011tunnelling}. The total tunneling probability is $\Gamma_{\psi} \sim e^{ -2 \mbox{ Im}\, \left( \mathcal \chi_{ out} - \mathcal \chi_{ in} \right)}$, with 
\begin{equation}
\mathcal \chi_{ out}(r)= - \mathcal \chi_{ in} (r) = \int \mathrm{d} r \,\frac{\omega}{\sqrt{\tilde{A}(r)\tilde{B}(r)}}\:.
\end{equation}%
It is important to mention that, under the dominant energy condition and the Einstein equations, the functions \(\tilde{A}(r)\) and \(\tilde{B}(r)\) possess identical zeroes. Consequently, in the vicinity of \(r_H\), to first order, we have: $
\tilde{A}(r)\tilde{B}(r)=\tilde{A}^{'}(r_{ h})\tilde{B}^{'}(r_{ h})(r-r_{ h})^2+...$
and it is evident that a simple pole exists with a well--defined coefficient. By applying Feynman's technique, we get
\ie
2\mbox{ Im}\;\left( \mathcal \chi_{ out} - \mathcal \chi_{ in} \right) =\mbox{ Im}\int \mathrm{d} r \,\frac{4\omega}{\sqrt{\tilde{A}(r)\tilde{B}(r)}}=\frac{2\pi\omega}{\kappa},
\fe
where the surface gravity is given by \(\kappa= \frac{1}{2}\sqrt{\tilde{A}^{'}(r_{h})\tilde{B}^{'}(r_{h})}\). In this context, the expression \(\Gamma_{\psi} \sim e^{-\frac{2 \pi \omega}{\kappa}}\) leads to the following particle density, $n_{\psi}$, to our black hole solution 
\ie
n_{\psi} = \frac{\Gamma_{\psi}}{1+\Gamma_{\psi}}  = \frac{1}{e^{8 \pi  \omega  \left(\frac{3 \Theta ^2}{64 M}+M\right)}+1}.
\fe
In Fig. \ref{particledendityfermions}, we present the behavior of \(n_{\psi}\) for various values of \(\Theta\). Additionally, we compare our results with the standard Schwarzschild case.

\section{The evaporation process}

\begin{figure}
    \centering
      \includegraphics[scale=0.45]{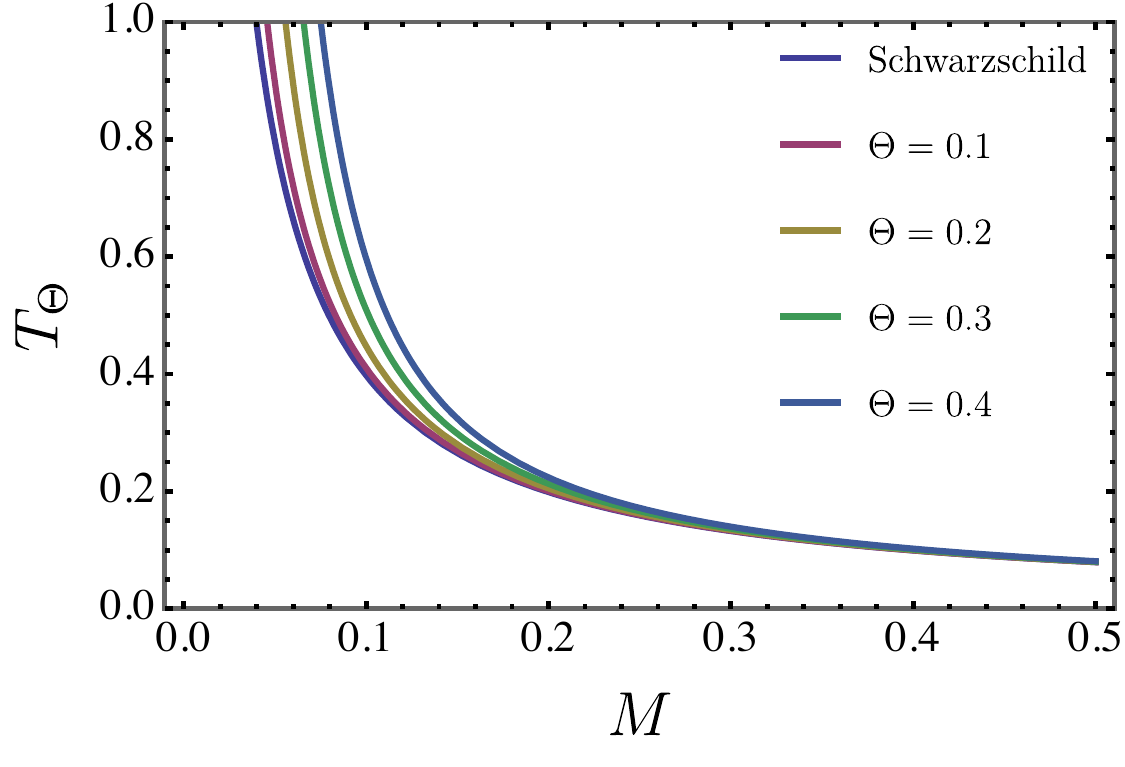}
    \caption{The \textit{Hawking} temperature $T_{\Theta}$ as a function of mass $M$ for different values of $\Theta$.}
    \label{hawkingtemperature}
\end{figure}

As shown in previous section, the \textit{Hawking} temperature is given by
\begin{equation}\label{Temp} 
\nonumber
 T_{\Theta} = \frac{1}{8 \pi  M}{+\frac{\Theta ^2}{256 \pi  M^3}}.
\end{equation}
In Fig. \ref{hawkingtemperature}, we represent the \textit{Hawking} temperature $T_{\Theta}$ as function of mass $M$ for different values of $\Theta$. As we can verify from the plots, the correction ascribed to the non--commutativity is small, as we should expect.
Notice that, in our case, the final stage of black hole evaporation ($T_\Theta \to 0$) leads to a {non--physical } remnant mass, i.e., $M_{rem}$, written below
\ie
\label{renmantmass111}
M_{rem} = {\frac{i \Theta }{4 \sqrt{2}}}.
\fe
This outcome contrasts with recent findings in the literature, where a different approach to non--commutativity was employed{, i.e., the Lorentzian distribution \cite{nascimento2024effects,araujo2025properties,filho2025particleasd}.}

In order to estimate the lifetime of the black hole, the calculation of the photon sphere as well the shadows is mandatory. To do so, we derive the computation of the photon sphere and shadows in a more general spherically symmetric scenario, as outlined follows. Initially, we consider a generic metric below
\ie
g_{\mu \nu}\mathrm{d}x^\mu \mathrm{d}x^\nu  =  - A(r)\mathrm{d}t^2 + B(r)\mathrm{d}{r^2} + C(r)\mathrm{d}\theta ^2 + D(r){{\mathop{\rm \sin}\nolimits} ^2}\theta \mathrm{d}\varphi ^2.
\fe
Now let us consider the Lagrangian method, we have 
\ie
\mathcal{L} = \frac{1}{2}{g_{\mu \nu }}{{\dot x}^\mu }{{\dot x}^\nu },
\fe
which reads
\begin{equation}
\mathcal{L} = \frac{1}{2}[ - A(r){{\dot t}^2} + B(r){{\dot r}^2} + C(r){{\dot \theta }^2} + D(r){{\mathop{\rm \sin}\nolimits} ^2}\, \theta {{\dot \varphi }^2}].
\end{equation}
By using the Euler--Lagrange equation and regarding the equatorial plane, i.e., $\theta = \frac{\pi }{2}$, we shall have two constant of motion $E$ and $L$ denoted as follows
\begin{equation}\label{constant}
E = A(r)\dot t \quad\mathrm{and}\quad L = D(r)\dot \varphi,
\end{equation}
For the light, therefore, it is written
\begin{equation}\label{light}
- A(r){{\dot t}^2} + B(r){{\dot r}^2} + D(r){{\dot \varphi }^2} = 0,
\end{equation}
After accomplishing some algebraic manipulations in putting Eq. (\ref{constant}) in Eq. (\ref{light}), we obtain
\begin{equation}\label{rdot}
\frac{{{{\dot r}^2}}}{{{{\dot \varphi }^2}}} = {\left(\frac{{\mathrm{d}r}}{{\mathrm{d}\varphi }}\right)^2} = \frac{{D(r)}}{{B(r)}}\left(\frac{{D(r)}}{{A(r)}}\frac{{{E^2}}}{{{L^2}}} - 1\right).
\end{equation}
Here, it is fundamental to notice that
\ie
\frac{\mathrm{d}r}{\mathrm{d}\lambda} = \frac{\mathrm{d}r}{\mathrm{d}\varphi} \frac{\mathrm{d}\varphi}{\mathrm{d}\lambda}  = \frac{\mathrm{d}r}{\mathrm{d}\varphi}\frac{L}{D(r)}, 
\fe
so that

\ie
\Dot{r}^{2} = \left( \frac{\mathrm{d}r}{\mathrm{d}\lambda} \right)^{2} =\left( \frac{\mathrm{d}r}{\mathrm{d}\varphi} \right)^{2} \frac{L^{2}}{D(r)^{2}}.
\fe
After given these preliminaries, let us particularize to our case. In this sense, we write $A(r) = 1-\frac{2 M_{\Theta}}{r}$, $B(r) =  \left( 1-\frac{2 M_{\Theta}}{r}\right)^{-1}$, $C(r) = r^{2}$ and $D(r) = r^{2}\sin^{2}\theta$. Therefore, it is straightforward to express
\ie
\Dot{r}^{2} = E^{2} + \mathcal{V}(r,\Theta),
\fe
where $\mathcal{V}(r,\Theta)$ is given by
\ie
\mathcal{V}(r,\Theta) = \frac{L^2 (2 M_{\Theta}-r)}{r^3}.
\fe
For the sake of obtaining the light sphere, we need to solve $\mathrm{d}\mathcal{V}/\mathrm{d}r = 0$. Remarkably, the solution for this equation gives rise to seven different solutions. Nevertheless, only one of them turns out to be a physical solution, $r_{c}$ as shown as follows
\ie
\begin{split}
\label{photonsphere}
r_{c} = 3 M_{\Theta}.
\end{split}
\fe
Now, let us derive the expression to the shadow radii as well
\ie
\mathcal{R} = \left.  \sqrt{\frac{D(r)}{A(r)}}   \right|_{r = {r_{c}}} = 3 \sqrt{3} M_{\Theta}.
\fe

\begin{figure}
    \centering
      \includegraphics[scale=0.6]{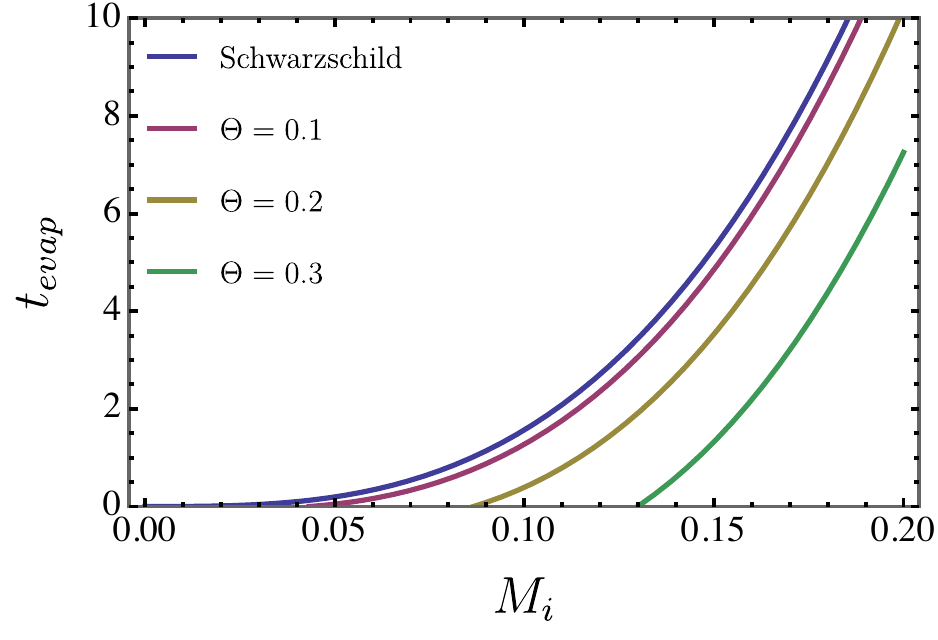}
    \caption{The behavior of the evaporation time $t_{evap}$ function of $M_{i}$ for different values of $\Theta$
.}
    \label{evaporationlifetime}
\end{figure}

Another significant feature that warrants investigation is the black hole's lifetime. To explore this, we write
\ie
\frac{\mathrm{d}M}{\mathrm{d}\tau} = - \alpha \sigma a T_{\Theta}^{4},
\fe
where \(a\) denotes the radiation constant, \(\sigma\) represents the cross--sectional area, and \(\alpha\) is the greybody factor. Under the geometric optics approximation, \(\sigma\) corresponds to the photon capture cross section
\ie
\sigma = \pi \left. \left( \frac{D(r)}{A(r)} \right)  \right|_{r = {r_{c}}}= 27 \pi \left( {M - \frac{ \Theta ^2}{ 32 M}}\right)^2,
\fe
so that
\ie
\begin{split}
\frac{\mathrm{d}M}{\mathrm{d}\tau} = -27 \pi  \Upsilon \left({M - \frac{ \Theta ^2}{ 32 M}}\right)^2 \left(\frac{1}{8 \pi  M} {+ \frac{  \Theta ^2}{256 \pi  M^3}}\right)^4
\end{split}
\fe
with $\Upsilon = a \alpha$. In this manner, it yields 
\ie
\begin{split}
\int_{0}^{t_{\text{evap}}} \Upsilon \mathrm{d}\tau & = - \int_{M_{i}}^{M_{f}} 
\left[    -27 \pi \left({M - \frac{ \Theta ^2}{ 32 M}}\right)^2 \left(\frac{1}{8 \pi  M}{+ \frac{  \Theta ^2}{256 \pi  M^3}}\right)^4 \right]^{-1} \mathrm{d}M,
\end{split}
\fe
where $(M_{i}$ refers to the initial mass configuration, and \(t_{\text{evap}}\) denotes the time corresponding to the final phase of the evaporation process. Importantly, this integral can be solved analytically, as demonstrated below
\ie
\begin{split}
t_{evap}  = & \frac{1}{ {324} \Upsilon} \pi ^3 \left\{ { +3072 \Theta ^2 (M_{f} - M_{i})+\frac{64 \Theta ^8 (M_{f}-M_{i})}{\left(\Theta ^2+32 (M_{f}-M_{i})^2\right)^3}+\frac{48 \Theta ^4 (M_{f} - M_{i})}{32 (M_{f}-\text{Mi})^2-\Theta ^2} }  \right.\\
& \left.  { -16384 (M_{f}-M_{i})^3 + \frac{496 \Theta ^6 (M_{i}-M_{f})}{\left(\Theta ^2+32 (M_{f}-M_{i})^2\right)^2}+\frac{2088 \Theta ^4 (M_{f} - M_{i})}{\Theta ^2+32 (M_{f}-M_{i})^2}  }  \right. \\
& \left.  {54 \sqrt{2} \Theta ^3 \tanh ^{-1}\left(\frac{4 \sqrt{2} (M_{f} - M_{i})}{\Theta }\right)-639 \sqrt{2} \Theta ^3 \tan ^{-1}\left(\frac{4 \sqrt{2} (M_{f} - M_{i})}{\Theta }\right)}   \right\},
\end{split}
\fe
and{, since there is not a physical remanant mass, we shall consider that the balck hole under consideration, at its final state, will evaporate completly, i.e., } $\lim\limits_{M_{f} \to {0}} t_{evap}$, we obtain
\ie
t_{evap} = {\frac{256}{81 \Upsilon} \pi ^3 \left( 16 M_{i}^3-3 \Theta ^2 M_{i} \right)},
\fe
{up to the second order of $\Theta$.} Moreover, taking into account $\lim\limits_{\Theta \to 0} t_{evap}$, we get
\ie
t_{evap_{\Theta=0}} = \frac{4096 \pi ^3 M_{i}^3}{81 \Upsilon },
\fe
{which corresponds exactly to the evaporation lifetime of a Schwarzschild black hole.} Now, let us see Fig. \ref{evaporationlifetime}, which displays the behavior of $t_{evap}$ as a function of $M_{i}$ for different values of $\Theta$. One important implication of our findings is that a usual Schwarzschild black hole possesses a finete life time. This indicates that the evaporation process will ultimately cease, resulting in {a possibility of} loss of information \cite{nicolini2006noncommutative,nozari2008hawking,perez2009thermodynamics}. 


\section{Conclusion}

In this work, we investigated a gravitationally non--commutative black hole by gauging the de Sitter SO(4,1) group and applying the Seiberg--Witten map. Our focus was on modifications of non--commutativity through mass deformation, as recently proposed in the literature \cite{heidari2023gravitational}.

Initially, we examined corrections to \textit{Hawking} radiation for bosonic particle modes by analyzing the Klein--Gordon equation in curved spacetimes. We computed the Bogoliubov coefficients, \(\alpha_{\omega\omega'}\) and \(\beta_{\omega\omega'}\), which indicated how \(\Theta\) introduced corrections to the amplitude associated with particle creation. We also derived the power spectrum \(P_{\Theta}(\omega)\) and the \textit{Hawking} temperature \(T_{\Theta}\) within this framework. Additionally, using the appropriate boundary conditions, we analyzed \textit{Hawking} radiation from a tunneling perspective, resulting in expressions for the power spectrum and particle number density based on the corrected emission rate \(\tilde{\Gamma} = e^{-8 \pi \omega M_{\Theta} \left( 1 - \frac{\omega}{2 M_{\Theta}} \right)}\). We also addressed the particle number density \(n\), finding that \(\Theta\) {increased} its magnitude.

A similar analysis was conducted for fermion particles, leading to a corresponding tunneling probability \(\Gamma_{\psi}\), which allowed us to determine the particle number density \(n_{\psi}\). In this case, \(\Theta\) was found to increase the magnitude of \(n_{\psi}\). The results were compared with those for the Schwarzschild case.

Furthermore, we found that, during the final stage of black hole evaporation, there {was no physical} residual mass \(M_{rem} = {\frac{i \Theta }{4 \sqrt{2}}}\). Notably, we derived an analytical expression for the black hole evaporation lifetime, given by \(t_{evap} = {\frac{256}{81 \Upsilon} \pi ^3 \left( 16 M_{i}^3-3 \Theta ^2 M_{i} \right)}\), and compared our findings with recent estimates of non--commutativity from the literature \cite{araujo2023thermodynamics}. Finally, as a future direction, we suggest extending similar investigations to other gravitational scenarios, such as bumblebee gravity within the metric and metric--affine formalisms. These and related ideas are currently under development.

In addition, as a natural next step one should extend our analysis beyond the $\mathcal{O}(\Theta^2)$ mass–deformation to include higher–order noncommutative corrections and thereby test the convergence of the semiclassical expansion as well as probe for any new, frequency–dependent deviations from thermality.  Another obvious avenue is to generalize the present construction to rotating and charged spacetimes so as to study how noncommutativity modifies superradiant scattering, charge‐loss rates, and the remnant mass formula in the Kerr– and Reissner–Nordström–like cases.  From the observational side, one may perform detailed ray–tracing in the deformed metric (including simple accretion-disc models) to assess the $\Theta$–induced shift of the photon‐ring diameter against EHT constraints, and calculate corrections to the quasinormal–mode spectrum for comparison with LIGO/Virgo ringdown data.  Finally, given the existence of a finite remnant mass, it would be of great interest to revisit the black–hole information puzzle—e.g.\ via the “island” prescription for entanglement entropy—in this noncommutative setting, and to carry out analogous mass–deformation programmes in other gauge–gravity frameworks such as bumblebee or metric–affine models to test the robustness of our conclusions.


\section*{Acknowledgements}

Most of the calculations were performed by using the \textit{Mathematica} software. Particularly, A. A. Araújo Filho is supported by Conselho Nacional de Desenvolvimento Científico e Tecnológico (CNPq) -- [150891/2023--7]. Particularly, A. A. Araújo Filho is grateful to S. M. Rodrigues for the valuable ideas and insightful discussions during the preparation of this manuscript. A.{\"O}. would like to acknowledge the contribution of the COST Action CA21106 - COSMIC WISPers in the Dark Universe: Theory, astrophysics and experiments (CosmicWISPers), the COST Action CA21136 - Addressing observational tensions in cosmology with systematics and fundamental physics (CosmoVerse), the COST Action CA22113 - Fundamental challenges in theoretical physics (THEORY-CHALLENGES), the COST Action CA23130 - Bridging high and low energies in search of quantum gravity (BridgeQG) and the COST Action CA23115 - Relativistic Quantum Information (RQI) funded by COST (European Cooperation in Science and
Technology). We also thank EMU, TUBITAK, ULAKBIM (Turkiye) and SCOAP3 (Switzerland) for their support.

\section{Data Availability Statement}

Data Availability Statement: No Data associated in the manuscript


\bibliographystyle{ieeetr}
\bibliography{ref}

\begin{thebibliography}{100}

\bibitem{wald2010general}
R.~M. Wald, {\em General relativity}.
\newblock University of Chicago press, 2010.

\bibitem{misner1973gravitation}
C.~W. Misner, K.~S. Thorne, and J.~A. Wheeler, {\em Gravitation}.
\newblock Macmillan, 1973.

\bibitem{szabo2003quantum}
R.~J. Szabo, ``Quantum field theory on noncommutative spaces,'' {\em Physics
  Reports}, vol.~378, no.~4, pp.~207--299, 2003.

\bibitem{szabo2006symmetry}
R.~J. Szabo, ``Symmetry, gravity and noncommutativity,'' {\em Classical and
  Quantum Gravity}, vol.~23, no.~22, p.~R199, 2006.

\bibitem{3}
N.~Seiberg and E.~Witten, ``String theory and noncommutative geometry,'' {\em
  Journal of High Energy Physics}, vol.~1999, no.~09, p.~032, 1999.

\bibitem{ferrari2004towards}
A.~F. Ferrari, H.~O. Girotti, M.~Gomes, A.~Y. Petrov, A.~Ribeiro, V.~O.
  Rivelles, and A.~da~Silva, ``Towards a consistent noncommutative
  supersymmetric yang-mills theory: Superfield covariant analysis,'' {\em
  Physical Review D}, vol.~70, no.~8, p.~085012, 2004.

\bibitem{ferrari2003finiteness}
A.~F. Ferrari, H.~O. Girotti, M.~Gomes, A.~Y. Petrov, A.~Ribeiro, and
  A.~Da~Silva, ``On the finiteness of noncommutative supersymmetric qed3 in the
  covariant superfield formulation,'' {\em Physics Letters B}, vol.~577,
  no.~1-2, pp.~83--92, 2003.

\bibitem{ferrari2004superfield}
A.~F. Ferrari, H.~O. Girotti, M.~Gomes, A.~Y. Petrov, A.~Ribeiro, V.~O.
  Rivelles, and A.~Da~Silva, ``Superfield covariant analysis of the divergence
  structure of noncommutative supersymmetric qed 4,'' {\em Physical Review D},
  vol.~69, no.~2, p.~025008, 2004.

\bibitem{chamseddine2001deforming}
A.~H. Chamseddine, ``Deforming einstein's gravity,'' {\em Physics Letters B},
  vol.~504, no.~1-2, pp.~33--37, 2001.

\bibitem{lopez2006towards}
J.~Lopez-Dominguez, O.~Obregon, M.~Sabido, and C.~Ramirez, ``Towards
  noncommutative quantum black holes,'' {\em Physical Review D}, vol.~74,
  no.~8, p.~084024, 2006.

\bibitem{nicolini2009noncommutative}
P.~Nicolini, ``Noncommutative black holes, the final appeal to quantum gravity:
  a review,'' {\em International Journal of Modern Physics A}, vol.~24, no.~07,
  pp.~1229--1308, 2009.

\bibitem{mann2011cosmological}
R.~B. Mann and P.~Nicolini, ``Cosmological production of noncommutative black
  holes,'' {\em Physical Review D}, vol.~84, no.~6, p.~064014, 2011.

\bibitem{modesto2010charged}
L.~Modesto and P.~Nicolini, ``Charged rotating noncommutative black holes,''
  {\em Physical Review D}, vol.~82, no.~10, p.~104035, 2010.

\bibitem{2}
G.~Zet, V.~Manta, and S.~Babeti, ``Desitter gauge theory of gravitation,'' {\em
  International Journal of Modern Physics C}, vol.~14, no.~01, pp.~41--48,
  2003.

\bibitem{1}
M.~Chaichian, A.~Tureanu, and G.~Zet, ``Corrections to schwarzschild solution
  in noncommutative gauge theory of gravity,'' {\em Physics Letters B},
  vol.~660, no.~5, pp.~573--578, 2008.

\bibitem{zhao2023quasinormal}
Y.~Zhao, Y.~Cai, S.~Das, G.~Lambiase, E.~Saridakis, and E.~Vagenas,
  ``Quasinormal modes in noncommutative schwarzschild black holes,'' {\em arXiv
  preprint arXiv:2301.09147}, 2023.

\bibitem{heidari2024exploring}
N.~Heidari, H.~Hassanabadi, A.~A. Araújo~Filho, and J.~Kriz, ``Exploring
  non-commutativity as a perturbation in the schwarzschild black hole:
  quasinormal modes, scattering, and shadows,'' {\em The European Physical
  Journal C}, vol.~84, no.~6, p.~566, 2024.

\bibitem{campos2022quasinormal}
J.~Campos, M.~Anacleto, F.~Brito, and E.~Passos, ``Quasinormal modes and shadow
  of noncommutative black hole,'' {\em Scientific Reports}, vol.~12, no.~1,
  p.~8516, 2022.

\bibitem{karimabadi2020non}
M.~Karimabadi, S.~A. Alavi, and D.~M. Yekta, ``Non-commutative effects on
  gravitational measurements,'' {\em Classical and Quantum Gravity}, vol.~37,
  no.~8, p.~085009, 2020.

\bibitem{araujo2023thermodynamics}
A.~A. Ara{\'u}jo~Filho, S.~Zare, P.~J. Porf{\'\i}rio, J.~K{\v{r}}{\'\i}{\v{z}},
  and H.~Hassanabadi, ``Thermodynamics and evaporation of a modified
  schwarzschild black hole in a non--commutative gauge theory,'' {\em Physics
  Letters B}, vol.~838, p.~137744, 2023.

\bibitem{myung2007thermodynamics}
Y.~S. Myung, Y.-W. Kim, and Y.-J. Park, ``Thermodynamics and evaporation of the
  noncommutative black hole,'' {\em Journal of High Energy Physics}, vol.~2007,
  no.~02, p.~012, 2007.

\bibitem{banerjee2008noncommutative}
R.~Banerjee, B.~R. Majhi, and S.~Samanta, ``Noncommutative black hole
  thermodynamics,'' {\em Physical Review D}, vol.~77, no.~12, p.~124035, 2008.

\bibitem{nozari2006reissner}
K.~Nozari and B.~Fazlpour, ``Reissner-nordstr$\backslash$"$\{$o$\}$ m black
  hole thermodynamics in noncommutative spaces,'' {\em arXiv preprint
  gr-qc/0608077}, 2006.

\bibitem{sharif2011thermodynamics}
M.~Sharif and W.~Javed, ``Thermodynamics of a bardeen black hole in
  noncommutative space,'' {\em Canadian Journal of Physics}, vol.~89, no.~10,
  pp.~1027--1033, 2011.

\bibitem{nozari2007thermodynamics}
K.~Nozari and B.~Fazlpour, ``Thermodynamics of noncommutative schwarzschild
  black hole,'' {\em Modern Physics Letters A}, vol.~22, no.~38,
  pp.~2917--2930, 2007.

\bibitem{araujo2023thermodynamical}
A.~A. Ara{\'u}jo~Filho, J.~Furtado, J.~A. A.~S. Reis, and J.~E.~G. Silva,
  ``Thermodynamical properties of an ideal gas in a traversable wormhole,''
  {\em Classical and Quantum Gravity}, vol.~40, no.~24, p.~245001, 2023.

\bibitem{furtado2023thermal}
J.~Furtado, H.~Hassanabadi, J.~Reis, {\em et~al.}, ``Thermal analysis of
  photon-like particles in rainbow gravity,'' {\em arXiv preprint
  arXiv:2305.08587}, 2023.

\bibitem{chaichian2008corrections}
M.~Chaichian, A.~Tureanu, and G.~Zet, ``Corrections to schwarzschild solution
  in noncommutative gauge theory of gravity,'' {\em Physics Letters B},
  vol.~660, no.~5, pp.~573--578, 2008.

\bibitem{nicolini2006noncommutative}
P.~Nicolini, A.~Smailagic, and E.~Spallucci, ``Noncommutative geometry inspired
  schwarzschild black hole,'' {\em Physics Letters B}, vol.~632, no.~4,
  pp.~547--551, 2006.

\bibitem{o1}
S.~W. Hawking, ``Particle creation by black holes,'' {\em Communications in
  mathematical physics}, vol.~43, no.~3, pp.~199--220, 1975.

\bibitem{o11}
S.~W. Hawking, ``Black hole explosions?,'' {\em Nature}, vol.~248, no.~5443,
  pp.~30--31, 1974.

\bibitem{o111}
S.~W. Hawking, ``Black holes and thermodynamics,'' {\em Physical Review D},
  vol.~13, no.~2, p.~191, 1976.

\bibitem{gibbons1977cosmological}
G.~W. Gibbons and S.~W. Hawking, ``Cosmological event horizons, thermodynamics,
  and particle creation,'' {\em Physical Review D}, vol.~15, no.~10, p.~2738,
  1977.

\bibitem{eeeKuang:2017sqa}
X.-M. Kuang, J.~Saavedra, and A.~\"Ovg\"un, ``{The Effect of the Gauss-Bonnet
  term to Hawking Radiation from arbitrary dimensional Black Brane},'' {\em
  Eur. Phys. J. C}, vol.~77, no.~9, p.~613, 2017.

\bibitem{eeeKuang:2018goo}
X.-M. Kuang, B.~Liu, and A.~\"Ovg\"un, ``{Nonlinear electrodynamics AdS black
  hole and related phenomena in the extended thermodynamics},'' {\em Eur. Phys.
  J. C}, vol.~78, no.~10, p.~840, 2018.

\bibitem{eeeOvgun:2015box}
A.~\"Ovg\"un and K.~Jusufi, ``{Massive vector particles tunneling from
  noncommutative charged black holes and their GUP-corrected thermodynamics},''
  {\em Eur. Phys. J. Plus}, vol.~131, no.~5, p.~177, 2016.

\bibitem{eeeOvgun:2019jdo}
A.~\"Ovg\"un, I.~Sakall\i{}, J.~Saavedra, and C.~Leiva, ``{Shadow cast of
  noncommutative black holes in Rastall gravity},'' {\em Mod. Phys. Lett. A},
  vol.~35, no.~20, p.~2050163, 2020.

\bibitem{eeeOvgun:2019ygw}
A.~\"Ovg\"un and I.~Sakall\i{}, ``{Hawking Radiation via Gauss-Bonnet
  Theorem},'' {\em Annals Phys.}, vol.~413, p.~168071, 2020.

\bibitem{o3}
B.~Harms and Y.~Leblanc, ``Statistical mechanics of black holes,'' {\em
  Physical Review D}, vol.~46, no.~6, p.~2334, 1992.

\bibitem{o4}
C.~Vaz, ``Canonical quantization and the statistical entropy of the
  schwarzschild black hole,'' {\em Physical Review D}, vol.~61, no.~6,
  p.~064017, 2000.

\bibitem{araujo2024dark}
A.~A. Ara{\'u}jo~Filho, K.~Jusufi, B.~Cuadros-Melgar, and G.~Leon, ``Dark
  matter signatures of black holes with yukawa potential,'' {\em Physics of the
  Dark Universe}, vol.~44, p.~101500, 2024.

\bibitem{o6}
D.~Hansen, D.~Kubiz{\v{n}}{\'a}k, and R.~B. Mann, ``Criticality and surface
  tension in rotating horizon thermodynamics,'' {\em Classical and Quantum
  Gravity}, vol.~33, no.~16, p.~165005, 2016.

\bibitem{o7}
D.~Chen, J.~Tao, {\em et~al.}, ``The modified first laws of thermodynamics of
  anti-de sitter and de sitter space--times,'' {\em Nuclear Physics B},
  vol.~918, pp.~115--128, 2017.

\bibitem{o8}
D.~Hansen, D.~Kubiz{\v{n}}{\'a}k, and R.~B. Mann, ``Universality of p- v
  criticality in horizon thermodynamics,'' {\em Journal of High Energy
  Physics}, vol.~2017, no.~1, pp.~1--24, 2017.

\bibitem{o9}
A.~Jawad and A.~Khawer, ``Thermodynamic consequences of well-known regular
  black holes under modified first law,'' {\em The European Physical Journal
  C}, vol.~78, pp.~1--10, 2018.

\bibitem{sedaghatnia2023thermodynamical}
P.~Sedaghatnia, H.~Hassanabadi, P.~J. Porf{\'\i}rio, W.~S. Chung, {\em et~al.},
  ``Thermodynamical properties of a deformed schwarzschild black hole via dunkl
  generalization,'' {\em arXiv preprint arXiv:2302.11460}, 2023.

\bibitem{araujo2023analysis}
A.~A. Ara{\'u}jo~Filho, ``Analysis of a regular black hole in verlinde’s
  gravity,'' {\em Classical and Quantum Gravity}, vol.~41, no.~1, p.~015003,
  2023.

\bibitem{aa2024implications}
A.~A. Ara{\'u}jo~Filho, ``Implications of a simpson--visser solution in
  verlinde’s framework,'' {\em The European Physical Journal C}, vol.~84,
  no.~1, pp.~1--22, 2024.

\bibitem{o10}
P.~Kraus and F.~Wilczek, ``Self-interaction correction to black hole
  radiance,'' {\em Nuclear Physics B}, vol.~433, no.~2, pp.~403--420, 1995.

\bibitem{011}
M.~K. Parikh and F.~Wilczek, ``Hawking radiation as tunneling,'' {\em Physical
  review letters}, vol.~85, no.~24, p.~5042, 2000.

\bibitem{o12}
M.~Parikh, ``A secret tunnel through the horizon,'' {\em International Journal
  of Modern Physics D}, vol.~13, no.~10, pp.~2351--2354, 2004.

\bibitem{o13}
M.~K. Parikh, ``Energy conservation and hawking radiation,'' {\em arXiv
  preprint hep-th/0402166}, 2004.

\bibitem{touati2024quantum}
A.~Touati and Z.~Slimane, ``Quantum tunneling from schwarzschild black hole in
  non-commutative gauge theory of gravity,'' {\em Physics Letters B}, vol.~848,
  p.~138335, 2024.

\bibitem{calmet2023quantum}
X.~Calmet, S.~D. Hsu, and M.~Sebastianutti, ``Quantum gravitational corrections
  to particle creation by black holes,'' {\em Physics Letters B}, vol.~841,
  p.~137820, 2023.

\bibitem{johnson2020hawking}
G.~Johnson and J.~March-Russell, ``Hawking radiation of extended objects,''
  {\em Journal of High Energy Physics}, vol.~2020, no.~4, pp.~1--16, 2020.

\bibitem{vanzo2011tunnelling}
L.~Vanzo, G.~Acquaviva, and R.~Di~Criscienzo, ``Tunnelling methods and
  hawking's radiation: achievements and prospects,'' {\em Classical and Quantum
  Gravity}, vol.~28, no.~18, p.~183001, 2011.

\bibitem{silva2013quantum}
C.~Silva and F.~Brito, ``Quantum tunneling radiation from self-dual black
  holes,'' {\em Physics Letters B}, vol.~725, no.~4-5, pp.~456--462, 2013.

\bibitem{anacleto2015quantum}
M.~Anacleto, F.~Brito, and E.~Passos, ``Quantum-corrected self-dual black hole
  entropy in tunneling formalism with gup,'' {\em Physics Letters B}, vol.~749,
  pp.~181--186, 2015.

\bibitem{mitra2007hawking}
P.~Mitra, ``Hawking temperature from tunnelling formalism,'' {\em Physics
  Letters B}, vol.~648, no.~2-3, pp.~240--242, 2007.

\bibitem{zhang2005new}
J.~Zhang and Z.~Zhao, ``New coordinates for kerr--newman black hole
  radiation,'' {\em Physics Letters B}, vol.~618, no.~1-4, pp.~14--22, 2005.

\bibitem{medved2002radiation}
A.~Medved, ``Radiation via tunneling from a de sitter cosmological horizon,''
  {\em Physical Review D}, vol.~66, no.~12, p.~124009, 2002.

\bibitem{del2024tunneling}
F.~Del~Porro, S.~Liberati, and M.~Schneider, ``Tunneling method for hawking
  quanta in analogue gravity,'' {\em arXiv preprint arXiv:2406.14603}, 2024.

\bibitem{mirekhtiary2024tunneling}
F.~Mirekhtiary, A.~Abbasi, K.~Hosseini, and F.~Tulucu, ``Tunneling of
  rotational black string with nonlinear electromagnetic fields,'' {\em Physica
  Scripta}, vol.~99, no.~3, p.~035005, 2024.

\bibitem{senjaya2024bocharova}
D.~Senjaya, ``The bocharova--bronnikov--melnikov--bekenstein black hole’s
  exact quasibound states and hawking radiation,'' {\em The European Physical
  Journal C}, vol.~84, no.~6, p.~607, 2024.

\bibitem{Ali:2025nrm}
R.~Ali, X.~Tiecheng, and R.~Babar, ``{Evaluation of physical properties of
  Kiselev like AdS spacetime in the context of f(R,T) gravity under the impact
  of quantum gravity},'' {\em Phys. Dark Univ.}, vol.~48, p.~101868, 2025.

\bibitem{Ali:2024wmj}
R.~Ali, X.~Tiecheng, and R.~Babar, ``{First-order quantum corrections of
  tunneling radiation in modified Schwarzschild\textendash{}Rindler black
  hole},'' {\em Gen. Rel. Grav.}, vol.~56, no.~2, p.~13, 2024.

\bibitem{Ali:2024tty}
R.~Ali, X.~Tiecheng, and R.~Babar, ``{Study of first-order quantum corrections
  of thermodynamics to a Dyonic black hole solution surrounded by a perfect
  fluid},'' {\em Nucl. Phys. B}, vol.~1008, p.~116710, 2024.

\bibitem{Ali:2022omt}
R.~Ali, R.~Babar, and M.~Asgher, ``{Gravitational Analysis of Rotating Charged
  Black-Hole-Like Solution in Einstein\textendash{}Gauss\textendash{}Bonnet
  Gravity},'' {\em Annalen Phys.}, vol.~534, no.~10, p.~2200074, 2022.

\bibitem{Ali:2021mtp}
R.~Ali and M.~Asgher, ``{Tunneling analysis under the influences of
  Einstein\textendash{}Gauss\textendash{}Bonnet black holes gravity theory},''
  {\em New Astron.}, vol.~93, p.~101759, 2022.

\bibitem{araujo2025particle}
A.~A. Ara{\'u}jo~Filho, ``Particle creation and evaporation in kalb-ramond
  gravity,'' {\em Journal of Cosmology and Astroparticle Physics}, vol.~2025,
  no.~04, p.~076, 2025.

\bibitem{araujo2025does}
A.~A. Ara{\'u}jo~Filho, ``How does non-metricity affect particle creation and
  evaporation in bumblebee gravity?,'' {\em arXiv e-prints}, pp.~arXiv--2501,
  2025.

\bibitem{filho2025particleasd}
A.~A. Ara{\'u}jo~Filho, ``Particle production induced by a lorentzian
  non--commutative spacetime,'' {\em arXiv preprint arXiv:2502.19366}, 2025.

\bibitem{heidari2023gravitational}
N.~Heidari, H.~Hassanabadi, A.~A. Ara{\'u}jo~Filho, J.~Kriz, S.~Zare, and P.~J.
  Porf{\'\i}rio, ``Gravitational signatures of a non--commutative stable black
  hole,'' {\em Physics of the Dark Universe}, p.~101382, 2023.

\bibitem{araujo2025neutrino}
A.~A. Ara{\'u}jo~Filho, N.~Heidari, and Y.~Shi, ``Neutrino dynamics in a
  non-commutative spacetime,'' {\em arXiv e-prints}, pp.~arXiv--2504, 2025.

\bibitem{AraujoFilho:2025viz}
A.~A. Ara\'ujo~Filho, N.~Heidari, and A.~\"Ovg\"un, ``{Axisymmetric black hole
  in a non-commutative gauge theory: classical and quantum gravity effects},''
  2 2025.

\bibitem{AraujoFilho:2024mvz}
A.~A. Ara\'ujo~Filho, N.~Heidari, and A.~\"Ovg\"un, ``{Geodesics, accretion
  disk, gravitational lensing, time delay, and effects on neutrinos induced by
  a non-commutative black hole},'' 12 2024.

\bibitem{Juric:2025kjl}
T.~Juri\'c, A.~N. Kumara, and F.~Po\v{z}ar, ``{Constructing noncommutative
  black holes},'' {\em Nucl. Phys. B}, vol.~1017, p.~116950, 2025.

\bibitem{9}
D.~Kastor, S.~Ray, and J.~Traschen, ``Enthalpy and the mechanics of ads black
  holes,'' {\em Classical and Quantum Gravity}, vol.~26, no.~19, p.~195011,
  2009.

\bibitem{hawking1975particle}
S.~W. Hawking, ``Particle creation by black holes,'' in {\em Euclidean quantum
  gravity}, pp.~167--188, World Scientific, 1975.

\bibitem{parker2009quantum}
L.~Parker and D.~Toms, {\em Quantum field theory in curved spacetime: quantized
  fields and gravity}.
\newblock Cambridge university press, 2009.

\bibitem{hollands2015quantum}
S.~Hollands and R.~M. Wald, ``Quantum fields in curved spacetime,'' {\em
  Physics Reports}, vol.~574, pp.~1--35, 2015.

\bibitem{wald1994quantum}
R.~M. Wald, {\em Quantum field theory in curved spacetime and black hole
  thermodynamics}.
\newblock University of Chicago press, 1994.

\bibitem{fulling1989aspects}
S.~A. Fulling, {\em Aspects of quantum field theory in curved spacetime}.
\newblock No.~17, Cambridge university press, 1989.

\bibitem{parikh2004energy}
M.~K. Parikh, ``Energy conservation and hawking radiation,'' {\em arXiv
  preprint hep-th/0402166}, 2004.

\bibitem{o69}
R.~Kerner and R.~B. Mann, ``Fermions tunnelling from black holes,'' {\em
  Classical and Quantum Gravity}, vol.~25, no.~9, p.~095014, 2008.

\bibitem{o70}
A.~Yale, ``Exact hawking radiation of scalars, fermions, and bosons using the
  tunneling method without back-reaction,'' {\em Physics Letters B}, vol.~697,
  no.~4, pp.~398--403, 2011.

\bibitem{o71}
R.~Di~Criscienzo and L.~Vanzo, ``Fermion tunneling from dynamical horizons,''
  {\em Europhysics Letters}, vol.~82, no.~6, p.~60001, 2008.

\bibitem{o72}
H.-L. Li, S.-Z. Yang, T.-J. Zhou, and R.~Lin, ``Fermion tunneling from a vaidya
  black hole,'' {\em Europhysics Letters}, vol.~84, no.~2, p.~20003, 2008.

\bibitem{o73}
R.~Kerner and R.~B. Mann, ``Charged fermions tunnelling from kerr--newman black
  holes,'' {\em Physics letters B}, vol.~665, no.~4, pp.~277--283, 2008.

\bibitem{o74}
A.~Yale and R.~B. Mann, ``Gravitinos tunneling from black holes,'' {\em Physics
  Letters B}, vol.~673, no.~2, pp.~168--172, 2009.

\bibitem{o75}
M.~Rehman and K.~Saifullah, ``Charged fermions tunneling from accelerating and
  rotating black holes,'' {\em Journal of Cosmology and Astroparticle Physics},
  vol.~2011, no.~03, p.~001, 2011.

\bibitem{o76}
B.~Chatterjee and P.~Mitra, ``Hawking temperature and higher order
  calculations,'' {\em Physics Letters B}, vol.~675, no.~2, pp.~240--242, 2009.

\bibitem{o77}
A.~Yale, ``There are no quantum corrections to the hawking temperature via
  tunneling from a fixed background,'' {\em The European Physical Journal C},
  vol.~71, pp.~1--4, 2011.

\bibitem{o83}
A.~Barducci, R.~Casalbuoni, and L.~Lusanna, ``Supersymmetries and the
  pseudoclassical relativistic electron,'' {\em Nuovo Cimento. A}, vol.~35,
  no.~3, pp.~377--399, 1976.

\bibitem{o84}
G.~Cognola, L.~Vanzo, S.~Zerbini, and R.~Soldati, ``On the lagrangian
  formulation of a charged spinning particle in an external electromagnetic
  field,'' {\em Physics Letters B}, vol.~104, no.~1, pp.~67--69, 1981.

\bibitem{vagnozzi2022horizon}
S.~Vagnozzi, R.~Roy, Y.-D. Tsai, L.~Visinelli, M.~Afrin, A.~Allahyari,
  P.~Bambhaniya, D.~Dey, S.~G. Ghosh, P.~S. Joshi, {\em et~al.},
  ``Horizon-scale tests of gravity theories and fundamental physics from the
  event horizon telescope image of sagittarius a,'' {\em Classical and Quantum
  Gravity}, 2022.

\bibitem{nascimento2024effects}
A.~A. Araújo~Filho, J.~R. Nascimento, A.~Y. Petrov, P.~J. Porf{\'\i}rio, and
  A.~{\"O}vg{\"u}n, ``Effects of non-commutative geometry on black hole
  properties,'' {\em Physics of the Dark Universe}, vol.~46, p.~101630, 2024.

\bibitem{araujo2025properties}
A.~A. Ara{\'u}jo~Filho, J.~R. Nascimento, A.~Y. Petrov, P.~J. Porf{\'\i}rio,
  and A.~{\"O}vg{\"u}n, ``Properties of an axisymmetric lorentzian
  non-commutative black hole,'' {\em Physics of the Dark Universe}, vol.~47,
  p.~101796, 2025.

\bibitem{nozari2008hawking}
K.~Nozari and S.~H. Mehdipour, ``Hawking radiation as quantum tunneling from a
  noncommutative schwarzschild black hole,'' {\em Classical and Quantum
  Gravity}, vol.~25, no.~17, p.~175015, 2008.

\bibitem{perez2009thermodynamics}
S.~P{\'e}rez-Pay{\'a}n and M.~Sabido, ``Thermodynamics of the schwarzschild
  black hole in noncommutative space,'' in {\em AIP Conference Proceedings},
  vol.~1116, pp.~451--453, American Institute of Physics, 2009.

\end{thebibliography}

\end{document}